\documentstyle[12pt]{article}

\newcommand{\be}{\begin{equation}}
\newcommand{\ee}{\end{equation}}
\newcommand{\Dlt}{\Delta}
\newcommand{\dlt}{\delta}
\newcommand{\prt}{\partial}
\newcommand{\br}{{\bf r}}
\newcommand{\bk}{{\bf k}}
\newcommand{\bt}{\beta}
\newcommand{\vp}{\varphi}

\newcommand{\al}{\alpha}
\newcommand{\ra}{\rightarrow}
\newcommand{\sgm}{\sigma}

\newcommand{\om}{\omega}
\newcommand{\Om}{\Omega}

\newcommand{\dgr}{\dagger}
\newcommand{\lbd}{\lambda}
\newcommand{\Lbd}{\Lambda}
\newcommand{\cF}{{\cal F}}

\begin{document}

\begin{center}

{\Large {\bf Nonequilibrium Bose systems and nonground-state Bose-Einstein 
condensates} \\ [5mm]

V.I. Yukalov$^{1,2}$} \\ [3mm]

{\it $^1$Bogolubov Laboratory of Theoretical Physics, \\
Joint Institute for Nuclear Research, Dubna 141980, Russia \\ [3mm]
$^2$ Theoretische Physik, Universit\"at Essen, \\
Universit\"atsstrasse 5, Essen 45117, Germany}

\end{center}

\begin{abstract}

The theory of resonant generation of nonground-state Bose-Einstein 
condensates is extended to Bose-condensed systems at finite temperature. 
The generalization is based on the notion of representative statistical 
ensembles for Bose systems with broken global gauge symmetry. 
Self-consistent equations are derived describing an arbitrary nonequilibrium 
nonuniform Bose system. The notion of finite-temperature topological coherent 
modes, coexisting with a cloud of noncondensed atoms, is introduced. It is 
shown that resonant generation of these modes is feasible for a gas of 
trapped Bose atoms at finite temperature.

\end{abstract}

\vskip 1cm


\vskip 1cm


\section{Introduction}

Statistical systems with Bose-Einstein condensate exhibit a variety of 
very interesting phenomena, both equilibrium as well as nonequilibrium 
(see, e.g., book [1] and review articles [2--5]). One such a nontrivial 
effect is the possibility of creating nonground-state Bose-Einstein 
condensates of trapped atoms, as was advanced in Ref. [6]. The 
properties of these nonground-state condensates have been studied 
in a number of papers, for instance in Refs. [6--19], and are recently 
summarized in survey [20]. All these works [6--20] deal solely with 
the case of trapped atoms forming a dilute gas at zero-temperature, 
when all atoms can be condensed, so that there is no noncondensed 
atoms. However the latter are always present in real experiments 
at finite temperature. Interactions between atoms also produce an 
admixture of noncondensed particles. Then the question arises whether 
the nonground-state condensates can be generated, when the trapped 
atoms do not form a purely condensed system, but there is also a cloud 
of noncondensed atoms.

To answer the above question, it is necessary to be based on a reliable 
general theory of nonequilibrium nonuniform Bose-condensed systems. The 
description of such systems is commonly done by employing the Bogolubov 
ideas [21--24], when the global gauge symmetry is broken by means of the 
Bogolubov shift. However, Hohenberg and Martin [25], first, emphasized
that all theories of Bose systems with broken gauge symmetry suffer 
from an internal inconsistency, being either nonconserving or displaying 
an unphysical gap in the collective spectrum. In both these cases, such 
theories do not correspond to stable statistical systems. A detailed 
discussion of this problem has recently been done by Andersen [3].

To overcome the standard deficiency of theories with broken gauge 
symmetry, it is necessary to invoke the notion of representative 
ensembles [26]. This notion is strictly formulated in Sec. 2 for 
arbitrary statistical systems. The appropriate representative ensemble 
for a Bose system with broken gauge symmetry is constructed in Sec. 3. 
The resulting evolution equations are derived from the general variational 
principle for the extremum of action functional, which makes all equations 
self-consistent and the theory conserving and gapless, as is shown in Sec. 
4. The derived evolution equation for the condensate wave function is 
analysed in Sec. 5, where topological coherent modes at finite temperature 
are defined. Resonant generation of these coherent modes, corresponding 
to nonground-state condensates, is demonstrated to be feasible even in 
the presence of a substantial admixture of noncondensed atoms.

Throughout the paper, the system of units is used, where the Planck and
Boltzmann constants are set to unity, $\hbar=1$, $k_B=1$.

\section{Representative statistical ensembles}

Statistical systems are characterized by statistical ensembles, which 
means the following. First, one has to specify a space $\cF$ of 
microstates spanning all admissible quantum or dynamical states of the 
given system. Second, a statistical operator $\hat\rho(0)$ at the initial 
time $t=0$ has to be fixed. Third, temporal evolution of either the 
statistical operator or of observable quantities is to be defined, which 
is given by defining the action of the time derivative $\prt/\prt t$. 
Thus, a general {\it nonequilibrium statistical ensemble} can be denoted 
as a triplet
\be
\label{1}
\left \{ \cF,\hat\rho(0),\frac{\prt}{\prt t}\right \} \; .
\ee

For stationary or, in particular, equilibrium statistical systems, 
when their evolution is either trivial or absent, an {\it equilibrium 
statistical ensemble} is a pair $\{\cF,\hat\rho\}$, where 
$\hat\rho(t)=\hat\rho(0)\equiv\hat\rho$.

The temporal evolution equation of operators in the Heisenberg 
representation can be symbolized with the help of the evolution operator 
$U(t)$, so that the evolution of an operator $\hat C(t)$ is given by the 
relation
$$
\hat C(t) = \hat U^+(t) \hat C(0) \hat U(t) \; .
$$
Then the {\it statistical average} of this operator is
$$
<\hat C(t)> \; \equiv \; {\rm Tr}\; \hat\rho(0) \hat C(t) = 
{\rm Tr}\; \hat\rho(t)\hat C(0) \; ,
$$
where the trace operator is accomplished over the given state of microstates
$\cF$.

Each statistical system is characterized by a set of dynamical variables. 
The latter, keeping in mind quantum statistical systems, can be called field 
variables, whose example is a set $\psi(x,t)=[\psi_j(x,t)]$ of field 
operators $\psi_j(x,t)$, with an index $j$ enumerating the set members. Here 
$t$ is time and $x$ is a collection of all other variables, which, e.g., could
be spatial coordinates or momenta.

The system energy operator is given by a Hamiltonian $\hat H[\psi]$, which 
is a functional of the field variables $\psi$. The related Lagrangian is
\be
\label{2}
\hat L[\psi] \equiv \int \psi^\dgr(x,t) \; i \; \frac{\prt}{\prt t}\;
\psi(x,t)\; dx \; - \; \hat H[\psi]\; ,
\ee
where the set $\psi=[\psi_j]$ can be treated as a column.

Strictly speaking, for correctly defining a statistical system, it is not 
always sufficient to fix just a Hamiltonian $\hat H[\psi]$ or a Lagrangian 
$\hat L[\psi]$, but it is necessary to formulate additional conditions or 
constraints for making the system uniquely defined. Suppose there is a family 
$\{\hat C_i[\psi]\}$ of self-adjoint operators $\hat C_i[\psi]=\hat 
C_i^+[\psi]$, which will be called {\it condition operators}, when the 
required additional constraints, imposed on the system, are formulated as 
{\it statistical conditions}
\be
\label{3}
C_i = \; <\hat C_i[\psi]> \; ,
\ee
defined as the statistical averages of the condition operators. A statistical 
ensemble can correctly represent the considered statistical system only when 
all appropriate statistical conditions are accurately taken into account.

{\it Representative statistical ensemble} is a statistical ensemble that 
correctly represents the given statistical system, uniquely defining all 
its physical properties. The construction of representative ensembles for 
equilibrium systems is described in Ref. [27]. Then the equilibrium statistical 
operators are obtained from the conditional minimization of information 
functionals [28]. Now, we shall generalize the definition of representative 
ensembles to arbitrary nonequilibrium statistical systems.

Let us consider a physical system, whose correct definition requires the 
validity of statistical conditions (3) for some condition operators. It 
is worth stressing that the latter are assumed to be self-adjoint, but they 
do not need to be compulsorily the integrals of motion. For instance, the 
normalization condition $N=<\hat N>$ for the total number of particles $N$ 
involves the number-of-particle operator $\hat N$, which does not commute 
with the Hamiltonian, when the global gauge symmetry is broken.

The most general way of deriving the evolution equations is by extremizing 
an action functional [29]. For a system under constraints, imposed by the 
statistical conditions (3), this implies the conditional extremization of 
the {\it effective action}
\be
\label{4}
A[\psi] \equiv \int \left\{ \hat L[\psi] - 
\sum_i \nu_i \hat C_i[\psi]\right \} \; dt \; ,
\ee
with the Lagrange multipliers $\nu_i$ guaranteeing the validity of constraints 
(3). Defining the {\it grand Hamiltonian}
\be
\label{5}
H[\psi] \equiv \hat H[\psi] +\sum_i \nu_i \hat C_i[\psi] \; ,
\ee
and using Lagrangian (2), we have for the action functional (4) the form
\be
\label{6}
A[\psi] = \int \left\{ \int \psi^\dgr(x,t)\; i \; 
\frac{\prt}{\prt t} \; \psi(x,t) - H[\psi]\right \}\; dt \; .
\ee

The extremization of the effective action, defined by the variation
\be
\label{7}
\dlt A[\psi] = 0 
\ee
with respect to the field variables $\psi$ and $\psi^\dgr$, implies the 
validity of the variational equations
\be
\label{8}
\frac{\dlt A[\psi]}{\dlt\psi^\dgr_j(x,t)} = 0 \; , \qquad
\frac{\dlt A[\psi]}{\dlt\psi_j(x,t)} = 0 \; .
\ee
In view of action (6), this is equivalent to the equations
\be
\label{9}
i\; \frac{\prt}{\prt t}\; \psi_j(x,t) = 
\frac{\dlt H[\psi]}{\dlt\psi^\dgr_j(x,t)}
\ee
and their Hermitian conjugate.

Thus, we obtain the evolution equations (9) for the field variables. As is 
seen, the evolution is governed by the grand Hamiltonian (5). In the Heisenberg 
representation, a field operator $\psi_j$ satisfies the Heisenberg equation of 
motion
$$
i\; \frac{\prt}{\prt t}\; \psi_j(x,t) = \left [ \psi_j(x,t), H[\psi]
\right ] \; ,
$$
which is equivalent to Eq. (9). Then the time evolution of the field operator
is described by the relation
$$
\psi_j(x,t) =\hat U^+(t)\psi_j(x,0)\hat U(t) \; ,
$$
with the evolution operator satisfying the Schr\"odinger equation
\be
\label{10}
i\; \frac{d}{dt}\; \hat U(t) = H[\psi(x,0)]\hat U(t) \; .
\ee

The evolution equations, either (9) or (10), is a necessary component 
for defining a nonequilibrium statistical ensemble. This ensemble is 
representative, since the evolution equations are derived with taking 
account of all statistical conditions uniquely characterizing the considered 
statistical system. The given additional constraints define the grand 
Hamiltonian (5) governing the system evolution. As is clear, the properties 
of a system under the given constraints can be essentially different from 
the properties of a system under other or without constraints. This is why 
the usage of representative ensembles is crucially important for correctly 
describing physical systems. The general nonequilibrium representative 
ensemble, in the case of an equilibrium system, reduces to the equilibrium 
representative ensemble with the same grand Hamiltonian (5).

\section{Broken gauge symmetry}

Now we shall use the notion of representative ensembles, formulated above, 
for developing a correct general theory for Bose systems with broken gauge 
symmetry, which is associated with the appearance of Bose-Einstein condensate.
The standard way of breaking the global gauge symmetry is by means of the 
Bogolubov shift
\be
\label{11}
\psi(\br,t) \longrightarrow \hat\psi(\br,t) \equiv 
\eta(\br,t) +\psi_1(\br,t) \; ,
\ee
in which $\eta(\br,t)$ is the condensate wave function and $\psi_1(\br,t)$ 
is the field operator of uncondensed atoms, with $\br$ being the spatial 
coordinate. The operators $\psi$ and $\psi_1$ satisfy the same Bose 
commutation relations. The passage from $\psi$ to $\psi_1$ corresponds to a 
canonical transformation realizing nonequivalent operator representations [30].
The Fock space $\cF(\psi)$, generated by the field operator $\psi^\dgr$,
characterizes the space of microstates for a system without Bose-Einstein 
condensate, while the Fock space $\cF(\psi_1)$, generated by the field operator 
$\psi_1^\dgr$, is the space of microstates for a system with Bose-Einstein 
condensate. The spaces $\cF(\psi)$ and $\cF(\psi_1)$ are mutually orthogonal 
[30]. The condensate wave function $\eta(\br,t)$ is the same as the coherent 
field. The field variables of condensed and uncondensed atoms are mutually 
orthogonal, such that
\be
\label{12}
\int \eta^*(\br,t)\psi_1(\br,t)\; d\br = 0 \; .
\ee
The Bogolubov shift (11) is sufficient for breaking the global gauge symmetry. 
Note that the method of infinitesimal sources is not always able to break the 
gauge symmetry [26].

Having now two field variables, $\eta$ and $\psi_1$, instead of just 
one $\psi$, requires to have two normalization conditions. One is the 
normalization of the condensate wave function to the number of condensed atoms
\be
\label{13}
N_0 \equiv \int |\eta(\br,t)|^2 d\br \; ,
\ee
where $N_0$, in general, can be a function of time, which is not marked 
explicitly just for brevity. Defining the operator $\hat N_0\equiv N_0\hat 1$, 
with $\hat 1$ being a unity operator in $\cF(\psi_1)$, we can represent 
normalization (13) in the standard form of a statistical condition
\be
\label{14}
N_0 = \; <\hat N_0> \; ,
\ee
as in Eq. (3). Here and in what follows, all statistical averages are 
accomplished over the space of microstates $\cF(\psi_1)$.

The second normalization condition is that one for the number of uncondensed 
atoms
\be
\label{15}
N_1 = \; <\hat N_1> \; ,
\ee
where the operator for the number of uncondensed atoms is
\be
\label{16}
\hat N_1 \equiv \int \psi_1^\dgr(\br,t) \psi_1(\br,t)\; d\br \; .
\ee
In general, $N_1$ can also be a function of time in a nonequilibrium system.

The total number of atoms is
\be
\label{17}
N = N_0 + N_1 \; ,
\ee
which could be considered as a normalization condition, instead of Eq. 
(15). For two field variables, we have to fix not less and not more than two 
normalization conditions. 

The atomic densities and the related atomic fractions are denoted as
$$
\rho_0 \equiv \frac{N_0}{V} \; , \qquad \rho_1\equiv \frac{N_1}{V} \; , 
\qquad \rho \equiv \frac{N}{V} \; ,
$$
\be
\label{18}
n_0 \equiv \frac{N_0}{N} = \frac{\rho_0}{\rho} \; , \qquad 
n_1\equiv \frac{N_1}{N} = \frac{\rho_1}{\rho} \; ,
\ee
where $V$ is the system volume. For these, one has
\be
\label{19}
\rho=\rho_0 + \rho_1 \; , \qquad n_0+n_1 = 1 \; .
\ee
It is assumed that $\rho_0$ and $n_0$ are not zero in the thermodynamic 
limit.

The number of condensed atoms $N_0$ is defined so that to make the system 
stable. In equilibrium, $N_0$ is to be found from the minimization of a 
thermodynamic potential. For a nonequilibrium system, $N_0$ must ensure the 
dynamic stability of the solution for the condensate function $\eta(\br,t)$ 
satisfying the related evolution equation.

For a system with broken gauge symmetry, the average $<\psi_1>$ may be 
nonzero. This, however, would mean that quantum numbers, such as spin or 
momentum, are not conserved. One, therefore, has to impose an additional 
constraint
\be
\label{20}
<\psi_1(\br,t)>\; = \; 0 \; .
\ee
By introducing the condition operator
\be
\label{21}
\hat\Lbd[\hat\psi] \equiv \int \left [
\lbd(\br,t) \psi_1^\dgr(\br,t) + \lbd^*(\br,t)\psi_1(\br,t)
\right ] \; dt \; ,
\ee
in which $\lbd(\br,t)$ is a complex function, constraint (20) can be rewritten
in the standard form (3) as the {\it quantum-number conservation condition}
\be
\label{22}
<\hat \Lbd[\hat\psi]> \; = \; 0 \; .
\ee
The grand Hamiltonian (5) has to be defined by taking into account the 
statistical conditions (14), (15), and (22), which yields
\be
\label{23}
H[\eta,\psi_1] = \hat H[\hat\psi] - \mu_0\hat N_0 - \mu_1\hat N_1 -
\Lbd[\hat\psi] \; .
\ee
The action functional (6), with the Bogolubov shift (11), takes the form
\be
\label{24}
A[\eta,\psi_1] = \int \left\{ \int \left [ 
\eta^*(\br,t)\; i\; \frac{\prt}{\prt t}\; \eta(\br,t) +
\psi_1^\dgr(\br,t)\; i\; \frac{\prt}{\prt t}\; \psi_1(\br,t)\right ] \; d\br
- H[\eta,\psi_1]\right \} \; dt \; .
\ee
The extremization of the action functional (24) implies two variational 
equations, for the condensate function,
\be
\label{25}
\frac{\dlt A[\eta,\psi_1]}{\dlt\eta^*(\br,t)} = 0 \; ,
\ee
and for the field operator of uncondensed atoms,
\be
\label{26}
\frac{\dlt A[\eta,\psi_1]}{\dlt\psi^\dgr_1(\br,t)} = 0 \; .
\ee
These equations, in view of the action functional (24), are equivalent to 
the evolution equations
\be
\label{27}
i\; \frac{\prt}{\prt t}\; \eta(\br,t) = 
\frac{\dlt H[\eta,\psi_1]}{\dlt\eta^*(\br,t)}
\ee
and
\be
\label{28}
i\; \frac{\prt}{\prt t}\; \psi_1(\br,t) = 
\frac{\dlt H[\eta,\psi_1]}{\dlt\psi_1^\dgr(\br,t)} \; .
\ee

Let us take the energy operator in the standard form
$$
\hat H[\hat\psi] = \int \hat\psi^\dgr(\br,t) \left ( - \; 
\frac{\nabla^2}{2m} + U \right ) \hat\psi(\br,t)\; d\br +
$$
\be
\label{29}
+ \frac{1}{2}\; \int \hat\psi^\dgr(\br,t) \hat\psi^\dgr(\br',t)
\Phi(\br-\br')\hat\psi(\br',t)\hat\psi(\br,t)\; d\br d\br' \; ,
\ee
in which $U=U(\br,t)$ is an external field and $\Phi(\br)=\Phi(-\br)$ is 
an interaction potential. To satisfy the conservation condition (22), the 
grand Hamiltonian (23) must have no linear in $\psi_1$ terms [31]. This is 
achieved by chosing the Lagrange multiplier
\be
\label{30}
\lbd(\br,t) = \left [ - \; \frac{\nabla^2}{2m} + U +
\int \Phi(\br-\br') |\eta(\br',t)|^2 d\br' \right ] 
\eta(\br,t)  \; .
\ee
The evolution equations (27) and (28) yield 
$$
i\; \frac{\prt}{\prt t}\; \eta(\br,t) =  
\left ( - \; \frac{\nabla^2}{2m} + U  -\mu_0 \right )\eta(\br,t) +
$$
\be
\label{31}
+ \int \Phi(\br-\br')\left [ |\eta(\br',t)|^2 
\eta(\br)+\hat X(\br,\br')\right ] \; d\br' 
\ee
and, respectively,
$$
i\; \frac{\prt}{\prt t}\; \psi_1(\br,t) =  
\left ( - \; \frac{\nabla^2}{2m} + U  -\mu_1 \right )\psi_1(\br,t) +
$$
\be
\label{32}
+ \int \Phi(\br-\br')\left [ |\eta(\br')|^2 \psi_1(\br)+ 
\eta^*(\br')\eta(\br)\psi_1(\br') + \eta(\br')\eta(\br)\psi_1^\dgr(\br') 
+ \hat X(\br,\br')\right ] \; d\br' \; ,
\ee
where, for brevity, the time-dependence is not explicitly shown in the 
right-hand sides of these equations, and the correlation operator
\be
\label{33}
\hat X(\br,\br') \equiv \psi_1^\dgr(\br')\psi_1(\br')\eta(\br) +
\psi_1^\dgr(\br')\eta(\br')\psi_1(\br) + 
\eta^*(\br')\psi_1(\br')\psi_1(\br) +
\psi_1^\dgr(\br')\psi_1(\br')\psi_1(\br)
\ee
is introduced.

To obtain an equation for the condensate wave function, we need to average 
Eq. (31). For this purpose, let us define the normal density matrix
\be
\label{34}
\rho_1(\br,\br') \equiv \; < \psi_1^\dgr(\br')\psi_1(\br)> \; ,
\ee
the anomalous density matrix
\be
\label{35}
\sgm_1(\br,\br') \equiv \; < \psi_1(\br')\psi_1(\br)> \; ,
\ee
and the densities
$$
\rho_0(\br) \equiv |\eta(\br)|^2 \; , \qquad
\rho_1(\br)\equiv\rho_1(\br,\br) = \; <\psi_1^\dgr(\br)\psi_1(\br)> \; , 
$$  
\be
\label{36}
\sgm_1(\br)\equiv\sgm_1(\br,\br) = \; <\psi_1(\br)\psi_1(\br)> \; .
\ee
The total density of atoms is the sum 
\be
\label{37}
\rho(\br) = \rho_0(\br) +\rho_1(\br) 
\ee
of the condensate density $\rho_0(\br)$ and of the density $\rho_1(\br)$ of 
uncondensed atoms. Also, let us use the notation
\be
\label{38}
\xi(\br,\br') \equiv \; < \psi_1^\dgr(\br')\psi_1(\br')\psi_1(\br)> \; .
\ee
Then the average of the correlation operator (33) becomes
\be
\label{39}
<\hat X(\br,\br')> \; = \rho_1(\br')\eta(\br) +\rho_1(\br,\br')\eta(\br')
+ \sgm_1(\br,\br')\eta^*(\br') + \xi(\br,\br') \; .
\ee
Finally, averaging Eq. (31), we find the equation for the condensate function
$$
i\; \frac{\prt}{\prt t}\; \eta(\br,t) = \left ( - \; \frac{\nabla^2}{2m} +
U - \mu_0 \right )\eta(\br) +
$$
\be
\label{40}
+ \int \Phi(\br-\br')\left [ \rho(\br')\eta(\br) + 
\rho_1(\br,\br')\eta(\br') + \sgm_1(\br,\br')\eta^*(\br') + 
\xi(\br,\br') \right ] \; d\br' \; ,
\ee
where again, for brevity, the temporal dependence in the right-hand side is 
not explicitly shown. This is a general equation valid for an arbitrary 
nonequilibrium Bose-condensed system.

It is worth noting that, contrary to Eq. (31), the average of Eq. (32) is 
not defined because of the following. The grand Hamiltonian does not change 
being complimented by the term
$$
z[\psi_1] \equiv \int <\zeta(\br,t)\psi_1^\dgr(\br,t) +
\zeta^*(\br,t)\psi_1(\br,t)>d\br \; ,
$$
which is identically zero owing to constraint (20) for any complex function 
$\zeta(\br,t)$. However, the variation of such a term results in
$$
\frac{\dlt z[\psi_1]}{\dlt\psi_1^\dgr(\br,t)} = \zeta(\br,t) \; ,
$$
which, generally, is not zero. Therefore, replacing $H[\eta,\psi_1]$ by 
$H[\eta,\psi_1]+z[\psi_1]$ would lead to the appearance in Eq. (28) of an 
additional undefined term $\zeta(\br,t)$. Equation (32) must be used for 
defining the evolution of the correlation functions $<\psi_1^\dgr\psi_1>$ 
and $<\psi_1\psi_1>$. But for these functions, an additive term $\zeta$ does 
not play any role, since $<\zeta\psi_1^\dgr>=<\zeta\psi_1>=0$ due to 
constraint (20).

\section{Self-consistent equations}

The normal and anomalous density matrices (34) and (35), entering the 
evolution equation (40) for the condensate function, can be expressed 
through Green functions. One needs the normal and anomalous Green functions 
[32--34], which can be assembled in a matrix $G(12)=[G_{\al\bt}(12)]$ with 
the elements
$$
G_{11}(12) = - i<\hat T\psi_1(1)\psi_1^\dgr(2)> \; , \qquad
G_{12}(12) = - i<\hat T\psi_1(1)\psi_1(2)> \; , 
$$
\be
\label{41}
G_{21}(12) = - i<\hat T\psi_1^\dgr(1)\psi_1^\dgr(2)> \; , \qquad
G_{22}(12) = - i<\hat T\psi_1^\dgr(1)\psi_1(2)> \; ,
\ee
where $\hat T$ is the chronological operator. Here and in what follows, the 
shorthand notation is used, denoting the set $\{\br_j,t_j\}$ just by the 
number $j$.

We shall also need the combination 
$$
\Psi(123) \equiv \psi_1(1)\psi_1(2)\psi_1^\dgr(3) +
\eta(1)\psi_1(2)\psi_1^\dgr(3)+ \psi_1(1)\eta(2)\psi_1^\dgr(3) +
$$
\be
\label{42}
+ \psi_1(1)\psi_1(2)\eta^*(3) + \eta(1)\eta(2)\psi_1^\dgr(3) +
\eta(1)\psi_1(2)\eta^*(3) + \psi_1(1)\eta(2)\eta^*(3) \; .
\ee
Using this, we define the binary Green function, which is a matrix 
$B(1234)=[B_{\al\bt}(1234)]$ with the elements
$$
B_{11}(1234) = -<\hat T\Psi(123)\psi_1^\dgr(4)> \; , \qquad
B_{12}(1234) = -<\hat T\Psi(123)\psi_1(4)> \; ,
$$
\be
\label{43}
B_{21}(1234) = -<\hat T\Psi^+(123)\psi_1^\dgr(4)> \; , \qquad
B_{22}(1234) = -<\hat T\Psi^+(123)\psi_1(4)> \; .
\ee
We may note that
$$
\lim_{t_{12}\ra-0} <\hat T\Psi(122)>\; = \; \rho_1(2)\eta(1) +
\rho_1(12)\eta(2) +\sgm_1(12)\eta^*(2) +\xi(12) \; = \;
<\hat X(12)> \; ,
$$
where $t_{12}=t_1-t_2$.

Introducing the retarded interaction potential
\be
\label{44}
\Phi(12) \equiv \Phi(\br_1-\br_2) \dlt(t_{12}+0) \; ,
\ee
we define the self-energy $\Sigma(12)$ by the relation
\be
\label{45}
\int \Sigma(13) G(32) \; d(3) =  i \int \Phi(13) B(1332)\; d(3) \; .
\ee
Then, from Eq. (32), we obtain the matrix equation
\be
\label{46}
\left ( \hat\tau  i \; \frac{\prt}{\prt t_1} + \frac{\nabla_1^2}{2m}
\; - \; U(1) +\mu_1 \right ) G(12) - \int \Sigma(13) G(32)\; d(3) =
\dlt(12)\hat 1 \; ,
\ee
in which
\begin{eqnarray}
\nonumber
\hat\tau \equiv \left [\begin{array}{cc}
1 & 0 \\
0 & -1 \end{array} \right ] \; , \qquad
\hat 1 \equiv \left [ \begin{array}{cc}
1 & 0 \\
0 & 1 \end{array}\right ] \; .
\end{eqnarray}
Equation (46) for the Green function is to be complimented by equation (40) 
for the condensate function, which can be represented as
\be
\label{47}
i\; \frac{\prt}{\prt t_1}\; \eta(1) = \left ( -\; \frac{\nabla_1^2}{2m}
+ U(1) - \mu_0 \right ) \eta(1) + \int \Phi(12)\left [ \rho_0(2)\eta(1)
+ <\hat T\Psi(122)>\right ] d(2) \; ,
\ee
where $\rho_0(1)\equiv|\eta(1)|^2$.

Equations (46) and (47) is a self-consistent set of equations derived from 
the extremization of the action functional (24). Hence, these equations 
respect all conservation laws of the Hamiltonian, as is should be for the 
equations derived from a variational procedure [3,35]. It is important to 
stress that the self-consistency of the equations is ensured by two, generally 
different, Lagrange multipliers $\mu_0$ and $\mu_1$, playing the role of the 
chemical potentials for condensed and uncondensed atoms, respectively. The 
potential $\mu_0$ guarantees the normalization condition (14), hence, $\mu_0=
\mu_0(N_0)$. The normalization condition (15) defines $\mu_1=\mu_1(N_1)$. The 
number of condensed atoms is to be such that to provide the stability of the 
system of $N$ atoms, so that $N_0=N_0(N)$. Because of Eq. (17), $N_1=N-N_0$. 
Therefore, $\mu_0=\mu_0(N)$ and $\mu_1=\mu_1(N)$. But there is no necessity 
that $\mu_0$ be equal to $\mu_1$, though it may occasionally happen. Thus,
this happens in the Bogolubov approximation [21--24], which is valid for 
asymptotically weak interactions. However, for more elaborate approximations, 
this is not so. To illustrate this fact, let us consider an equilibrium 
system, when $\eta(\br,t)=\eta(\br)$ does not depend on time. Then Eq. (47) 
becomes
$$
\left ( -\; \frac{\nabla^2}{2m} + U\right )\eta(\br) +
$$
\be
\label{48}
+ \int \Phi(\br-\br')\left [ \rho(\br')\eta(\br) + 
\rho_1(\br,\br')\eta(\br') + \sgm_1(\br,\br')\eta^*(\br') 
+\xi(\br,\br')\right ]\; d\br' = \mu_0 \eta(\br) \; .
\ee
The grand thermodynamic potential is
\be
\label{49}
\Om = - T\ln {\rm Tr}\exp \left ( -\bt H[\eta,\psi_1]\right ) \; ,
\ee
and the free energy being
\be
\label{50}
F = - T\ln{\rm Tr} \exp\left\{ -\bt\left ( \hat H[\hat\psi]-
\hat\Lbd[\hat\psi]\right ) \right \} \; ,
\ee
where $\bt=T^{-1}$ is inverse temperature. For potentials (49) and (50),
we have the relation
\be
\label{51}
\Om = F - \mu_0 N_0 - \mu_1 N_1 \; .
\ee
The number of condensed atoms is defined by the stability condition implying 
the minimum of the grand potential (49),
\be
\label{52}
\frac{\prt\Om}{\prt N_0} = 0 \; , \qquad 
\frac{\prt^2\Om}{\prt N_0^2} > 0  \; .
\ee
More generally, the condensate wave function, assuring the system stability, 
is defined by the condition
\be
\label{53}
\frac{\dlt\Om}{\dlt\eta(\br)} = \; 
< \frac{\dlt H[\eta,\psi_1]}{\dlt\eta(\br)} > \; = \; 0 \; .
\ee
The latter, owing to Eq. (27), gives Eq. (48).

Since $N_0=N_0(N)$ and $N_1=N_1(N)$, it is the total number of atoms $N$ 
that can only be fixed for a system, which requires the relation
\be
\label{54}
\Om = F - \mu N \; ,
\ee
in which $\mu$ is the system chemical potential. Comparing Eqs. (51) and 
(54), immediately results in the definition of the {\it system chemical 
potential}
\be
\label{55}
\mu = n_0\mu_0 + n_1\mu_1 \; .
\ee
For the latter, one has the standard thermodynamic equations
$$
\mu =\left ( \frac{\prt F}{\prt N}\right )_{TV} \; , \qquad
N = - \left ( \frac{\prt\Om}{\prt\mu}\right )_{TV} \; .
$$
Let us emphasize that, since $N_0$ and $N_1$ are uniquely defined through 
the total number of atoms $N$, neither $N_0$ nor $N_1$ can be treated as 
independent thermodynamic variables, because of which, generally, 
$\mu_0\neq\mu_1$.

To give a more explicit demonstration that $\mu_0\neq\mu_1$, let us 
turn to a uniform system, when the density matrices $\rho_1(\br,\br')$, 
$\sgm_1(\br,\br')$, as well as $\xi(\br,\br')$, depend solely on the 
difference $\br-\br'$. Then the densities $\rho_1=\rho_1(\br,\br)$, 
$\sgm_1(\br,\br)$, and $\xi_1(\br,\br)$ are constants, together with the 
order parameter $\eta=\eta(\br)$. Also, normalization (14) yields 
$\eta=\sqrt{\rho_0}$. From Eq. (48), we find
\be
\label{56}
\mu_0 = \rho\Phi_0 + \int \left ( n_k + \sgm_k +
\frac{\xi_k}{\sqrt{\rho_0}} \right ) \Phi_k\; \frac{d\bk}{(2\pi)^3} \; ,
\ee
where $\Phi_k$, $n_k$, $\sgm_k$, and $\xi_k$ are the Fourier transforms of
the corresponding quantities $\Phi(\br)$, $\rho_1(\br,0)$, $\sgm_1(\br,0)$, 
and $\xi(\br,0)$, respectively. For the Fourier transforms 
$G_{\al\bt}(\bk,\om)$ of the Green functions $G_{\al\bt}(12)$, we find the 
former from Eq. (46). The poles of $G_{\al\bt}(\bk,\om)$ define the 
single-atom spectrum, which is gapless, provided that
\be
\label{57}
\mu_1 = \Sigma_{11}(0,0) - \Sigma_{12}(0,0) \; ,
\ee
where $\Sigma_{\al\bt}(\bk,\om)$ is the Fourier transform of the 
self-energy $\Sigma_{\al\bt}(12)$. Note that the gapless spectrum is a
necessary requirement for the existence of a stable system with 
Bose-Einstein condensate [30]. Equation (57) is the Hugenholtz-Pines 
relation that could be derived either from thermodynamic equations [33] or 
from the Ward identities with respect to the variation of gauge [24].

One often considers Bose-condensed systems in the contact-potential 
approximation, when
$$
\Phi(\br) = \Phi_0 \dlt(\br) \qquad \left ( \Phi_0 \equiv 
4\pi\; \frac{a_s}{m} \right ) \; ,
$$
with $a_s$ being a scattering length. Then $\Phi_k=\Phi_0$, and Eq. (56) 
gives
\be
\label{58}
\mu_0 = \left ( \rho +\rho_1 +\sgm_1 + \frac{\xi}{\sqrt{\rho_0}} \right )
\Phi_0 \; ,
\ee
where
$$
\rho_1 = \int n_k\; \frac{d\bk}{(2\pi)^3} \; , \qquad
\sgm_1 = \int \sgm_k\; \frac{d\bk}{(2\pi)^3} \; , \qquad
\xi = \int \xi_k \; \frac{d\bk}{(2\pi)^3} \; .
$$
Clearly, there is no such a general law that would require the identity
of $\mu_0$ and $\mu_1$.

To be even more specific, let us consider the Hartree-Fock-Bogolubov 
approximation, when $\xi(\br,\br')=0$. Then for Eq. (57), we find
\be
\label{59}
\mu_1 = \rho\Phi_0 + \int (n_k-\sgm_k) \Phi_k \;
\frac{d\bk}{(2\pi)^3} \; .
\ee
In the case of the contact potential, we obtain
\be
\label{60}
\mu_0 = (\rho+\rho_1 + \sgm_1)\Phi_0 \; , \qquad
\mu_1 = (\rho+\rho_1 - \sgm_1)\Phi_0 \; .
\ee
As is evident, $\mu_0\neq\mu_1$.

In the standard approach, without using the representative ensemble, one 
assumes that $\mu_0$ equals $\mu_1$, hence, according to Eq. (55), equals 
$\mu$. But then one comes to the explicit inconsistency in Eqs. (60). A 
common way of trying to treat this inconsistency is by neglecting the 
anomalous average $\sgm_1$, calling this the "Popov approximation". 
However, this unjustified trick has nothing to do with Popov, as is easy 
to infer from his original works [36--38]. Moreover, it is easy to show 
that at low temperatures the anomalous average $\sgm_1$ can be much larger 
than the normal average $\rho_1$, that is, neglecting $\sgm_1$ has nothing 
to do with a reasonable approximation [30,39].

By employing the representative ensemble, as is done in the present paper, 
we introduce two Lagrange multipliers, $\mu_0$ and $\mu_1$, for two 
normalization conditions (14) and (15). These multipliers are not obliged 
to be equal, but are to be such that to render the whole theory completely 
self-consistent.

\section{Topological coherent modes}

Equation (48), defining the condensate function $\eta(\br)$, corresponds to 
a stable equilibrium system, when condensing atoms pile down to the 
ground-state quantum level. This equation can be generalized for describing 
arbitrary stationary states $\eta_n(\br)$, enumerated by a quantum 
multi-index $n$, so that, for the contact potential, we have
$$
\left [ - \; \frac{\nabla^2}{2m} + U(\br) \right ] \eta_n(\br) +
$$
\be
\label{61}
+ \Phi_0\left \{ \left [ |\eta_n(\br)|^2 + 2\rho_1^{(n)}(\br)\right 
]\eta_n(\br) + \sgm_1^{(n)}(\br)\eta_n^*(\br) + \xi^{(n)}(\br)\right \}
= E_n \eta_n(\br) \; ,
\ee
where $\rho_1^{(n)}(\br)$, $\sgm_1^{(n)}(\br)$, and $\xi^{(n)}(\br)$ are 
the solutions to the self-consistent system of equations for $\rho_1(\br)$,
$\sgm_1(\br)$, and $\xi(\br)\equiv\xi(\br,\br)$, respectively, under the 
condition that $\eta(\br)$ is replaced by $\eta_n(\br)$. The ground-state 
energy level corresponds to
\be
\label{62}
E_0 \equiv \min_n E_n = \mu_0 \; ,
\ee
with $\eta_0(\br)\equiv\eta(\br)$. But, in general, there can exist a 
whole set of coherent states $\eta_n(\br)$, with a spectrum of energies 
$E_n$. The condensate functions $\eta_n(\br)$ will be called the {\it 
topological coherent modes}.

It is clear that, for a system at absolute equilibrium, the sole pertinent 
mode is the ground state $\eta(\br)$. In order to generate excited 
coherent modes, it is necessary to deal with a nonequilibrium system. Then 
one has to consider the time-dependent equation (40) for the condensate 
function, which, for the contact potential, reads as 
$$
i\; \frac{\prt}{\prt t}\; \eta(\br,t) = \left ( - \; \frac{\nabla^2}{2m}
+ U - \mu_0 \right ) \eta(\br,t) +
$$
\be
\label{63}
+ \Phi_0\left\{ \left [ \rho_0(\br,t) + 2\rho_1(\br,t) \right ] 
\eta(\br,t) + \sgm_1(\br,t)\eta^*(\br,t) +\xi(\br,t) \right \}\; ,
\ee
where $U=U(\br,t)$ is a time-dependent external potential. The latter can 
be represented as a sum
\be
\label{64}
U(\br,t) = U(\br) + V(\br,t)
\ee
of a trapping potential $U(\br)$ and a modulating potential $V(\br,t)$.

If in Eq. (63) we set $\rho_1(\br,t)=0$, $\sgm_1(\br,t)=0$, and 
$\xi(\br,t)=0$, then we come to the usual temporal Gross-Pitaevskii 
equation [1]. The generation of various topological coherent modes on
the basis of the latter equation has been investigated earlier [6--20,40]. 
However, in general, we have to analyse the full Eqs. (61) and (63).

Suppose that at the initial time $t=0$, the atomic system is in 
equilibrium, so that
\be
\label{65}
\eta(\br,0) =\eta_0(\br) \equiv\eta(\br) \; .
\ee
Assume that we wish to generate a nonground-state condensate, corresponding 
to a topological coherent mode labelled by the index $n=n_1$, with the 
energy $E_1\equiv E_{n_1}$. For this purpose, we impose an alternating 
modulating 
potential
\be
\label{66}
V(\br,t) = V_1(\br)\cos\om t + V_2(\br)\sin \om t \; ,
\ee
with the frequency $\om$ tuned close to the transition frequency 
$\om_1\equiv E_1-E_0$, which implies the resonance condition
\be
\label{67}
\left | \frac{\Dlt\om}{\om} \right | \ll 1 \qquad
(\Dlt\om \equiv \om -\om_1) \; .
\ee

Imposing an external alternating field will, of course, destroy the 
ground-state condensate by means of two processes. One is the resonant 
process of transferring condensed atoms from the ground state mode, with 
the energy $E_0=\mu_0$, to the chosen excited mode, with the energy $E_1$. 
And also, there will be nonresonant processes of taking atoms to other 
nonresonant coherent modes, as well as the process of transferring atoms 
from the condensate to the cloud of noncondensed atoms. Estimates show 
[15] that it is feasible to arrange such a setup that the resonant 
generation would occur much faster than other nonresonant processes. 
Assuming this, we may consider the solution of Eq. (63) at times shorter 
than the critical time $t_c$, when the resonant generation prevails, and 
the number of condensed atoms stays practically constant,
\be
\label{68}
\int |\eta(\br,t)|^2 d\br = \int |\eta(\br)|^2 d\br = N_0 \; .
\ee
The topological coherent modes are also normalized to $N_0$, because of 
which we can set
\be
\label{69}
\eta_n(\br) =\sqrt{N_0}\; \vp_n(\br) \; , \qquad
\int |\vp_n(\br)|^2 d\br = 1 \; .
\ee
Then we look for the solution of Eq. (63) in the form of the mode 
expansion
\be
\label{70}
\eta(\br,t) =\sum_n C_n(t) \eta_n(\br) e^{-i\om_n t} \; ,
\ee
in which $\om_n\equiv E_n-E_0$ and the coefficient functions $C_n(t)$ are 
assumed to be slow functions of time, as compared to the exponential 
$\exp(-i\om_n t)$, which means the condition
\be
\label{71}
\frac{1}{\om_n}\left | \frac{dC_n}{dt}\right | \ll 1 \; .
\ee
From the normalization condition (68), with expansion (70), after 
averaging over time, while keeping the quasi-invariants $C_n(t)$ fixed,
we get
\be
\label{72}
\sum_n |C_n(t)|^2  = 1 \; .
\ee

Substituting expansion (70) into Eq. (63), we multiply the latter by 
$\eta_n^*(\br)\exp(i\om_nt)$, integrate over $\br$, and average over time, 
following the averaging technique [6,14,15]. To this end, we need the 
following integrals
$$
\al_{mn} \equiv \Phi_0 N_0 \int |\vp_m(\br)|^2 \left [
2|\vp_n(\br)|^2 - |\vp_m(\br)|^2 \right ]\; d\br \; ,
$$
$$
\bt_{mn} \equiv \int \vp_m^*(\br)\left [ V_1(\br) - i V_2(\br)\right ] 
\vp_n(\br) \; d\br \; ,
$$
$$
\overline\al_{nn}(t) \equiv \al_{nn} - \Phi_0 \int \vp_n^*(\br) \left\{
2\left [ \rho_1^{(n)}(\br) - \rho_1(\br,t)\right ] \vp_n(\br) +
\sgm_1^{(n)}(\br)\vp_n^*(\br) + \frac{\xi^{(n)}(\br)}{\sqrt{N_0}}
\right \} \; d\br \; ,
$$
$$
\al_{nn} = \Phi_0 N_0 \int |\vp_n(\br)|^4 d\br \; , \qquad
\Dlt_{mn} \equiv \Dlt\om +\al_{00} - \al_{11} \; .
$$
We presume that, at the resonant stage, when $t\ll t_c$, the quantity 
$\overline\al_{nn}(t)$ is a slow function of time, such that
\be
\label{73}
\frac{t_c}{|\overline\al_{nn}|} \left | \frac{d\overline\al_{nn}}{dt}
\right | \ll 1 \; .
\ee
Then, introducing the functions
\be
\label{74}
c_n(t) \equiv C_n(t)\exp\left\{ i\overline\al_{nn}(t) t\right \} \; ,
\ee
we, finally, arrive at the system of two equations
$$
i \; \frac{dc_0}{dt} = \al_{01}|c_1|^2 c_0 + \frac{1}{2}\; \bt_{01} c_1
\exp(i\Dlt_{01}t) \; ,
$$
\be
\label{75}
i \; \frac{dc_0}{dt} = \al_{10}|c_0|^2 c_1 + \frac{1}{2}\; \bt_{01}^* c_0
\exp(-i\Dlt_{01}t) \; .
\ee
Equations (75) have the same form as studied in the previous publications 
[6,8,10,14--16,20] describing the resonant generation of topological coherent 
modes on the basis of the Gross-Pitaevskii equation.

The reduction of the general equation for the condensate function (63) to 
the system of Eqs. (75) demonstrates that the topological coherent modes, 
corresponding to nonground-state condensates, can be resonantly generated 
also in the presence of the cloud of uncondensed atoms. The procedure of 
the resonant generation requires the validity of several conditions 
discussed above. In particular, the resonant process must be much faster 
than nonresonant ones. After the critical time $t_c$, power broadening 
destroys the resonant procedure and nonresonant processes become 
prevailing. According to estimates [15], the critical time is of the order 
of $t_c\sim \om/(\al^2+\bt^2)$, where $\al\equiv(\al_{01}+\al_{10})/2$ and
$\bt\equiv|\bt_{01}|$. At the initial stage of time $t\ll t_c$, the 
resonant generation is feasible. The critical time $t_c$ can be made 
comparable to the lifetime of atoms in a trap [15].


\begin{thebibliography}{99}
\bibitem{1}
L. Pitaevskii and S. Stringari, Bose-Einstein Condensation in Dilute Gases 
(Clarendon, Oxford, 2003).

\bibitem{2}
P.W. Courteille, V.S. Bagnato, and V.I. Yukalov, Laser Phys. {\bf 11}, 
659 (2001).

\bibitem{3}
J.O. Andersen, Rev. Mod. Phys. {\bf 76}, 599 (2004).

\bibitem{4}
K. Bongs and K. Sengstock, Rep. Prog. Phys. {\bf 67}, 907 (2004).

\bibitem{5}
V.I. Yukalov, Laser Phys. Lett. {\bf 1}, 435 (2004).

\bibitem{6}
V.I. Yukalov, E.P. Yukalova, and V.S. Bagnato, Phys. Rev. A {\bf 56}, 
4845 (1997).

\bibitem{7}
K.P. Marzlin and W. Zhang, Phys. Rev. A {\bf 57}, 4761 (1998).

\bibitem{6}
V.I. Yukalov, E.P. Yukalova, and V.S. Bagnato, Laser Phys. {\bf 10},
26 (2000).

\bibitem{9}
Y.S. Kivshar, T.J. Alexander, and S.K. Turitsyn, Phys. Lett. A {\bf 278}, 
225 (2001).

\bibitem{10}
V.I. Yukalov, E.P. Yukalova, and V.S. Bagnato, Laser Phys. {\bf 11},
455 (2001).

\bibitem{11}
R. D'Agosta, B.A. Malomed, and C. Presilla, Laser Phys. {\bf 12}, 37 
(2002).

\bibitem{12}
R. D'Agosta and C. Presilla, Phys. Rev. A {\bf 65}, 043609 (2002).

\bibitem{13}
N.P. Proukakis and L. Lambropoulos, Eur. Phys. J. D {\bf 19}, 355 (2002).

\bibitem{14}
V.I. Yukalov and E.P. Yukalova, J. Phys. A {\bf 35}, 8603 
(2002).

\bibitem{15}
V.I. Yukalov, E.P. Yukalova, and V.S. Bagnato, Phys. Rev. A {\bf 66},
043602 (2002).

\bibitem{16}
V.I. Yukalov, E.P. Yukalova, and V.S. Bagnato, Laser Phys. {\bf 13},
861 (2003).

\bibitem{17}
S.K. Adhikari, Phys. Lett. A {\bf 308}, 302 (2003).

\bibitem{18}
V.I. Yukalov, K.P. Marzlin, and E.P. Yukalova, Phys. Rev. A {\bf 69}, 
023620 (2004).

\bibitem{19}
S.K. Adhikari, Phys.Rev. A {\bf 69}, 063613 (2004).

\bibitem{20}
V.I. Yukalov, K.P. Marzlin, E.P. Yukalova, and V.S. Bagnato, Am. Inst. 
Phys. Conf. Proc. {\bf 770}, 218 (2005).

\bibitem{21}
N.N. Bogolubov, J. Phys. (Moscow) {\bf 11}, 23 (1947).

\bibitem{22}
N.N. Bogolubov, Moscow Univ. Phys. Bull. {\bf 7}, 43 (1947).

\bibitem{23}
N.N. Bogolubov, Lectures on Quantum Statistics (Gordon and
Breach, New York, 1967), Vol. 1.

\bibitem{24}
N.N. Bogolubov, Lectures on Quantum Statistics (Gordon and
Breach, New York, 1970), Vol. 2.

\bibitem{25}
P.C. Hohenberg and P.C. Martin, Ann. Phys. {\bf 34}, 291 (1965).

\bibitem{26}
V.I. Yukalov, Phys. Rev. E {\bf 72}, 066119 (2005).

\bibitem{27}
V.I. Yukalov, Phys. Rep. {\bf 208}, 395 (1991).

\bibitem{28}
V.I. Yukalov, Statistical Green's Functions (Queen's University, Kingston, 
1998).

\bibitem{29}
H. Kleinert, Path Integrals (World Scientific, Singapore, 2004).

\bibitem{30}
V.I. Yukalov, Laser Phys. {\bf 16}, 511 (2006).

\bibitem{31}
N.N. Bogolubov and N.N. Bogolubov Jr., Introduction to Quantum Statistical 
Mechanics (Gordon and Breach, Lausanne, 1994).

\bibitem{32}
S.T. Beliaev, J. Exp. Theor. Phys. {\bf 7}, 289 (1958).

\bibitem{33}
N.M. Hugenholtz and D. Pines, Phys. Rev. {\bf 116}, 489 (1959).

\bibitem{34}
P.O. Fedichev and G.V. Shlyapnikov, Phys. Rev. A {\bf 58}, 3146 (1998).

\bibitem{35}
G. Baym and G. Grinstein, Phys. Rev. D {\bf 15}, 2897 (1977).

\bibitem{36}
V.N. Popov, J. Exp. Theor. Phys. {\bf 20}, 1185 (1965).

\bibitem{37}
V.N. Popov, Functional Integrals in Quantum Field Theory and Statistical 
Physics (Reidel, Dordrecht, 1983).

\bibitem{38}
V.N. Popov, Functional Integrals and Collective Modes (Cambridge 
University, New York, 1987).

\bibitem{39}
V.I. Yukalov and E.P. Yukalova, Laser Phys. Lett. {\bf 2}, 506 
(2005).

\bibitem{40}
C. Weiss and T. Jinasundra, Phys. Rev. A {\bf 72}, 053626 (2005).


\end{thebibliography}
\end{document}